\documentclass[journal=jacsat,manuscript=article]{achemso}
\usepackage[version=3]{mhchem} 

\usepackage[T1]{fontenc} 
\usepackage{upgreek}
\usepackage{graphicx}
\usepackage[]{epsfig}
\usepackage{times,amsmath,amssymb}

\usepackage{color}
\usepackage[colorlinks = true,
            linkcolor = blue,
            urlcolor  = blue,
            citecolor = blue,
            anchorcolor = blue]{hyperref}

\author{Koichi Baba}
\affiliation{Research Institute of Electrical Communication, Tohoku University, 2-1-1 Katahira, Aoba-ku, Sendai 980-8577, Japan}
\alsoaffiliation{Graduate School of Engineering, Tohoku University, 6-6 Aramaki Aza Aoba, Aoba-ku, Sendai 980-0845, Japan}

\author{Kosuke Noro}
\affiliation{Research Institute of Electrical Communication, Tohoku University, 2-1-1 Katahira, Aoba-ku, Sendai 980-8577, Japan}
\alsoaffiliation{Graduate School of Engineering, Tohoku University, 6-6 Aramaki Aza Aoba, Aoba-ku, Sendai 980-0845, Japan}

\author{Yusuke Kozuka}
\affiliation{Research Center for Materials Nanoarchitectonics (MANA), National Institute for Material Science (NIMS),
1-1 Namiki, Tsukuba 305-0044, Japan}

\author{Takeshi Kumasaka}
\affiliation{Research Institute of Electrical Communication, Tohoku University, 2-1-1 Katahira, Aoba-ku, Sendai 980-8577, Japan}

\author{Motoya Shinozaki}
\affiliation{WPI Advanced Institute for Materials Research, Tohoku University, 2-1-1 Katahira, Aoba-ku, Sendai 980-8577, Japan}

\author{Masashi Kawasaki}
\affiliation{Department of Applied Physics and Quantum-Phase Electronics Center (QPEC), University of Tokyo, 7-3-1 Hongo, Bunkyo-ku, Tokyo 113-8656, Japan}
\alsoaffiliation{RIKEN Center for Emergent Matter Science (CEMS), 2-1 Hirosawa, Wako, Saitama 351-0198, Japan}

\author{Tomohiro Otsuka}
\email{tomohiro.otsuka@tohoku.ac.jp}
\affiliation{WPI Advanced Institute for Materials Research, Tohoku University, 2-1-1 Katahira, Aoba-ku, Sendai 980-8577, Japan}
\alsoaffiliation{Research Institute of Electrical Communication, Tohoku University, 2-1-1 Katahira, Aoba-ku, Sendai 980-8577, Japan}
\alsoaffiliation{Graduate School of Engineering, Tohoku University, 6-6 Aramaki Aza Aoba, Aoba-ku, Sendai 980-0845, Japan}
\alsoaffiliation{Center for Science and Innovation in Spintronics, Tohoku University, 2-1-1 Katahira, Aoba-ku, Sendai 980-8577, Japan}
\alsoaffiliation{RIKEN Center for Emergent Matter Science (CEMS), 2-1 Hirosawa, Wako, Saitama 351-0198, Japan}

\title{Formation of multiple quantum dots in ZnO heterostructures}


\begin{document}

\begin{abstract}
In recent years, advancements in semiconductor manufacturing technology have enabled the formation of high-quality, high-mobility two-dimensional electron gases in zinc oxide (ZnO) heterostructures, making the electrostatic formation of quantum dots possible. 
ZnO, with its low natural abundance of isotopes possessing nuclear spin and its direct bandgap, is considered a potentially suitable material for quantum bit applications. 
In this study, we achieve the formation of triple quantum dots and the realization of a few-electron state in ZnO heterostructure devices. 
We also confirm that by varying the gate voltage between the quantum dots, it is possible to control the interdot spacing. 
Additionally, we observe a tunneling phenomenon called a quantum cellular automata effect, where multiple electrons move simultaneously, which is not seen in single or double quantum dots, due to Coulomb interactions. Our results demonstrate that ZnO nanostructures have reached a level where they can function as controllable multiple quantum dot systems.
\end{abstract}

\section{}

\begin{figure*}
\begin{center}
  \includegraphics{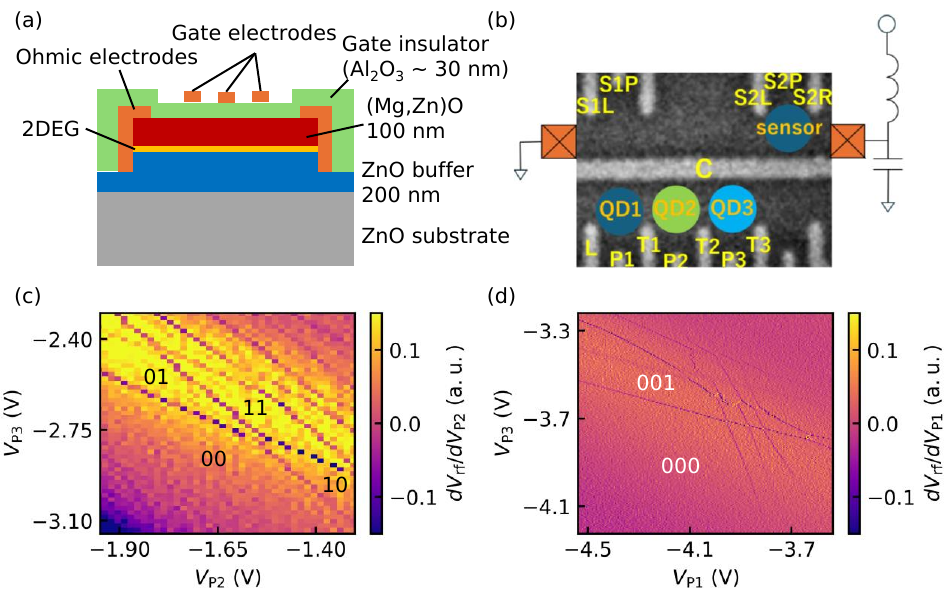}
  \caption{(a) A cross-sectional schematic view of the ZnO heterostructure device used in the experiment. (b) Scanning electron microscope image of the device. The positions of the formed quantum dots are also drawn. (c) Charge stability diagram of double quantum dots. The number of electrons in each quantum dot is denoted as (n$_{1}$ n$_{2}$). (d) Charge stability diagram of triple quantum dots. The number of electrons in each quantum dot is shown as (n$_{1}$ n$_{2}$ n$_{3}$).}
  \label{fig1}
\end{center}
\end{figure*}

Advancements in semiconductor microfabrication technology have made it possible to create nanoscale fine structures. 
A representative example of this is semiconductor quantum dots, which can confine electrons within nanoscale regions, enabling the observation of quantum mechanical properties~\cite{Tarucha1996prl,Ciorga2000prb,kouwenhoven2001few}. 
Research is underway to utilize the electron spins confined in quantum dots as quantum bits (qubits)~\cite{ono2002current,elzerman2004single,hanson2005single,amasha2008electrical,morello2010single}, the fundamental units of quantum information systems. 
Semiconductor quantum dots have the intrinsic potential for large-scale integration, and the spins of electrons confined within the dots exhibit long coherence times~\cite{petta2005coherent, koppens2006driven, yoneda2014fast, yoneda2018quantum, noiri2022fast, philips2022universal}, allowing for highly precise qubit operations. 
This has raised expectations for realizing quantum computers using semiconductor qubits~\cite{loss1998quantum,ladd2010quantum}.

Recently, high-quality and high-mobility two-dimensional electron gases (2DEGs) have been realized in zinc oxide (ZnO) heterostructure devices~\cite{tsukazaki2007quantum}. 
This has enabled the observation of quantum phenomena such as the quantum Hall effect~\cite{tsukazaki2007quantum, tsukazaki2010observation, falson2015even, falson2018review} and quantum point contacts~\cite{hou2019quantized} on ZnO devices, as well as the electrostatic formation of quantum dots~\cite{noro2024parity,noro2025charge}. 
ZnO, being a direct band gap semiconductor, exhibits strong coupling with light~\cite{Nakano_2008}, and its low abundance of isotopes with nuclear spin is expected to result in a long electron spin coherence time~\cite{Linpeng2018coherence,Niaouris2022ensemble}. 
These characteristics make ZnO a promising material for applications such as quantum bits.

To realize quantum computers, it is necessary to integrate quantum bits, and semiconductor qubits offer the advantage of utilizing existing integration technologies. 
Semiconductor spin qubits have already demonstrated high-precision quantum state manipulation~\cite{takeda2016fault, yoneda2018quantum, mkadzik2022precision, xue2022quantum, takeda2022quantum}, and their scalability has been advanced through the development of multi-quantum dot devices~\cite{takakura2014single, otsuka2016single,ito2016detection, ito2018four, philips2022universal, zwerver2022qubits, neyens2024probing}. 
However, research has thus far been limited to the formation of single and double quantum dots in ZnO heterostructures~\cite{noro2024parity, noro2025charge}, making the development of multiple quantum dots the next challenge for utilizing the unique properties of ZnO in quantum applications.

The scaling up of quantum dot systems contributes to quantum information processing with large-scale qubits while also providing platforms to explore fundamental quantum physics phenomena.
Quantum computers require integrating a large number of qubits, making the development of scalable multi-quantum dot architectures essential. 
Beyond applications, multi-quantum dot systems with three or more quantum dots exhibit unique physical phenomena not observable in single or double dots~\cite{busl2013bipolar, braakman2013long, sanchez2014long, sanchez2014super}.
Among these phenomena, the quantum cellular automata (QCA) effect~\cite{lent1993quantum, Islamshah1999science, gaudreau2006stability, schroer2007electrostatically}, a tunneling phenomenon where multiple electrons move simultaneously through Coulomb interactions, is of special interest.
This effect has been studied for information transfer between qubits or as a mechanism for quantum information processing itself. 
Experimental measurements of the real-time dynamics of this effect~\cite{aizawa2024dynamics} have provided fundamental insights into electron transport mechanisms in multi-quantum dot systems.

To utilize the advantages of ZnO, such as potential long spin coherence times, establishing multi-quantum dots with precise control capabilities is essential.
The formation of few-electron states and the gate voltage control of interdot couplings are fundamental requirements for both qubit operations and the investigation of quantum phenomena. 
Precise gate control enables the systematic exploration of quantum states and transport phenomena by allowing us to adjust energy levels and tunnel barriers.
These technological developments are the key to opening the potential of ZnO in both quantum information applications and fundamental quantum understanding.

In this study, we fabricate a multi-quantum dot sample using a ZnO heterostructure and obtain the charge stability diagram of a triple quantum dot. 
Furthermore, we demonstrate the formation of a few-electron state in the triple quantum dot and show that control between quantum dots is possible by varying the gate voltage. 
Additionally, we observe a correlated tunneling phenomenon in ZnO, where multiple electrons move simultaneously through Coulomb interactions, a phenomenon not observed in single or double quantum dots.

Figure~\ref{fig1}(a) shows a cross-sectional view of the ZnO heterostructure device used in this experiment. 
From top to bottom, it consists of a stacked structure of an Al$_{2}{\rm O}_{3}$ insulating film, (Mg, Zn)O, a ZnO buffer, and a ZnO substrate. 
Additionally, a 2DEG is formed at the interface between the (Mg, Zn)O layer and the ZnO buffer layer.

In this study, radio-frequency (rf) reflectometry is used to obtain charge stability diagrams. 
RF reflectometry~\cite{reilly2007fast, Barthel2010prb, kurzmann2019charge, noro2025charge} utilizes quantum point contacts or quantum dots as charge sensors, reading the resistance of the charge sensor as the intensity of rf reflections. 
Compared to transmission current measurements, it offers the advantages of faster and lower-noise measurements.
Figure~\ref{fig1}(b) shows a scanning electron micrograph of the fabricated device. 
Triple quantum dots are formed in the lower part, and the charge state is probed by a sensor quantum dot on the upper side. 
The Target QD is formed by L, P1, P2, P3, T1, T2, and T3. 
Note that the two gates in the lower-right corner are not used in this experiment. 
The quantum dots of the sensor are adjusted to form S2R, S2P, and S2L, constructing a resonance circuit as shown in the diagram. 
A resonant frequency of 174 MHz is used. 
All measurements are conducted in a dilution refrigerator with a base temperature of 60 mK.

Figure~\ref{fig1}(c) shows a charge stability diagram obtained by using P2, P3, T1, T2, and T3 to form double quantum dots at the positions of QD2 and QD3 in Fig.~\ref{fig1}(b) before forming the triple quantum dots, with settings $V_{\rm{C}} =-3.80$~V, $V_{\rm{L}} = -1.15$~V, $V_{\rm{T1}} = -2.00$~V, $V_{\rm{T2}} = -3.10$~V, $V_{\rm{T3}} = -3.40$~V, $V_{\rm{S2L}} = -3.34$~V, $V_{\rm{S2P}} = -3.25$~V, $V_{\rm{S2R}} = -3.38$~V. 
The reflected rf signal $V_{\mathrm{rf}}$ from the sensor quantum dot is measured while varying P2 and P3. 
The existence of two different slopes of the charge transition lines confirms the formation of double quantum dots. 
In Fig.~\ref{fig1}(c), we find that charge transition lines disappear over a wide range in the lower left region, where no electrons are present in each quantum dot. 
This result indicates that it is possible to realize and observe a few-electron state in our device.

Next, we form the triple quantum dots. 
Figure~\ref{fig1}(d) shows a charge stability diagram of triple quantum dots formed by applying voltages to L and P1 while keeping the double quantum dot state shown in Fig.~\ref{fig1}(c). 
With settings $V_{\rm{C}} =-3.80$~V, $V_{\rm{L}} = -1.33$~V, $V_{\rm{T1}} = -1.71$~V, $V_{\rm{T2}} = -3.20$~V, $V_{\rm{T3}} = -3.60$~V, $V_{\rm{P2}} = -1.93$~V, $V_{\rm{S2L}} = -3.34$~V, $V_{\rm{S2P}} = -3.34$~V, $V_{\rm{S2R}} = -3.38$~V, we measure the $V_{\mathrm{rf}}$ while varying P1 and P3. 
In Fig.~\ref{fig1}(d), the presence of three lines with different slopes corresponding to QD1, QD2, and QD3 confirms the formation of triple quantum dots. 
Additionally, similar to the charge stability diagram of the double quantum dots discussed earlier, the disappearance of charge transition lines over a wide range in the lower left region indicates that no electrons are present in each quantum dot in this area.

\begin{figure}
  \centering
  \includegraphics{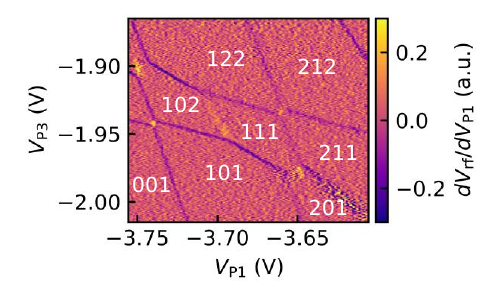}
  \caption{An enlarged view around the (111) regime in the charge stability diagram of the triple quantum dots in Fig.~\ref{fig1}(d).}
  \label{fig2}
\end{figure}

Figure~\ref{fig2} shows an enlarged view of the charge stability diagram of the triple quantum dot around the state (111). 
For quantum bit applications, the electron state in (111), in which each dot contains a single electron, is important. In Fig.~\ref{fig2}, $V_{\mathrm{rf}}$ are measured while changing P1 and P3 with $V_{\rm{C}} =-3.80$~V,$V_{\rm{L}} =-1.33$~V, $V_{\rm{T1}} = -1.71$~V, $V_{\rm{T2}} = -3.21$~V, $V_{\rm{T3}} = -3.60$~V, $V_{\rm{P2}} = -1.96$~V, $V_{\rm{S2L}}=-3.34$~V, $V_{\rm{S2P}} = -3.34$~V, $V_{\rm{S2R}} = -3.38$~V. 
We can observe a few electron charge states with electron numbers (n$_{1}$ n$_{2}$ n$_{3}$) indicated in the figure. 
The occurrence of gaps at the crossings of the charge transition lines indicates the presence of electrostatic coupling between the dots. 
The size of the gap depends on the capacitance between the quantum dots~\cite{hanson2007spins}. 
Therefore, the larger gap size suggests that the electrostatic coupling between QD1 and QD2, as well as between QD2 and QD3, is stronger than the coupling between QD1 and QD3.

\begin{figure*}
  \centering
  \includegraphics{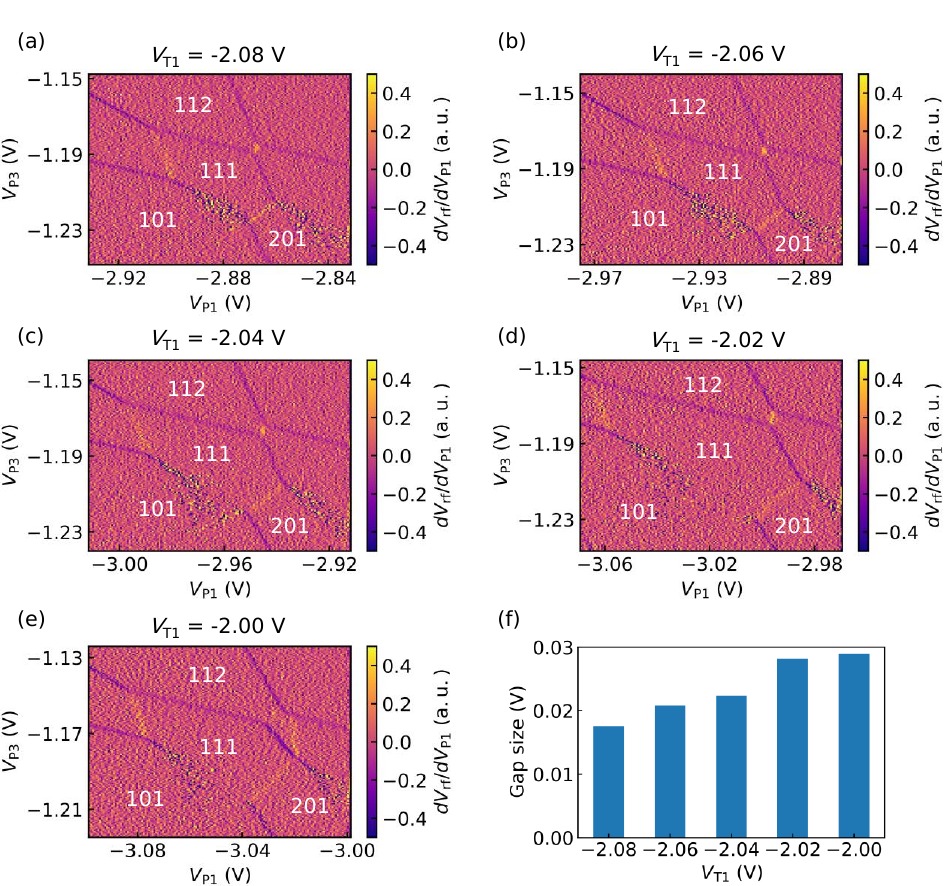}
  \caption{Tuning of the interdot coupling between QD1 and QD2; charge stability diagrams showing the voltage applied to T1 at different levels.  (a) $V_{\rm{T1}} = -2.08$~V. (b) $V_{\rm{T1}} = -2.06$~V. (c) $V_{\rm{T1}} = -2.04$~V. (d) $V_{\rm{T1}} = -2.02$~V. (e) $V_{\rm{T1}} = -2.00$~V. (f) The change in the size of the energy gap of level crossing in response to changes in T1.}
  \label{fig3}
\end{figure*}

Tuning of the interdot coupling is essential for controlling the electronic states and applications for qubits in the gate-defined multiple quantum dots.
By varying the gate voltage of T1, we can manipulate the coupling between QD1 and QD2 as shown in Figs.~\ref{fig3}(a)-(e). The voltages of the other electrodes are adjusted to keep the charge state around the (111) state.
Figure~\ref{fig3}(f) shows the gap size at the crossing, which corresponds to the length of the charge transition line between (111) and (201), as a function of the gate voltage of T1. 
This figure indicates that increasing the T1 voltage enhances the gap at the level crossing between the electron configurations (111) and (201) due to increased electrostatic coupling between QD1 and QD2. 
This result demonstrates that we can systematically control the coupling between quantum dots.

\begin{figure}
  \centering
  \includegraphics{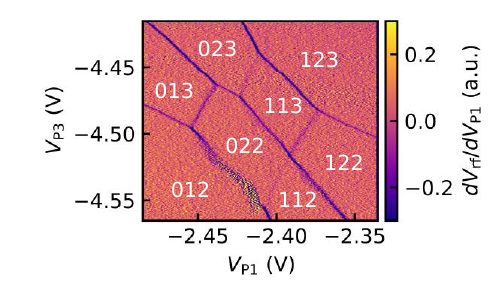}
  \caption{Charge stability diagram showing the QCA effect.}
  \label{fig4}

\end{figure}

Having demonstrated the triple quantum dots operation, we are ready to observe characteristic phenomena that are observed only in triple or more quantum dots.
We detect the tunneling phenomenon in which multiple electrons move simultaneously due to the Coulomb interaction QCA effect.
Figure~\ref{fig4} shows the QCA effect observed in the charge stability diagram of triple quantum dots. 
In this measurement, we sweep $V_{\rm{P1}}$ and $V_{\rm{P3}}$ with fixing the following gate voltages applied to each gate electrode: $V_{\rm{C}} =-3.50$~V, $V_{\rm{L}} = -1.63$~V, $V_{\rm{T1}} = -1.68$~V, $V_{\rm{T2}} = -2.93$~V, $V_{\rm{T3}} = -3.40$~V, $V_{\rm{P2}} = -1.17$~V.
These gate voltages are adjusted to cross the three kinds of charge transition lines corresponding to the three dots at a point and opening the gap. 
During rf reflectometry measurements, $V_{\rm{S1L}} = -2.35$~V, $V_{\rm{S1P}} = -1.72$~V are applied, and a quantum point contact is used as a sensor in this measurement with a probe frequency of approximately 220 MHz. 
By identifying the electron configurations from the charge transition lines, regions corresponding to electron numbers (012), (111), (022), (112), (013), (113), (122), and (123) are identified. 
In particular, the regions with electron configurations (022) and (113) are adjacent to each other. 
The transitions in this region require simultaneous movement of multiple electrons, and the QCA effect, which emerges only in more than three quantum dots, is observed in this region.

In this study, we have formed multiple quantum dots in a zinc oxide heterostructure device and demonstrated few-electron states in triple quantum dots.
We have shown that the interdot coupling can be systematically controlled by varying the gate voltage between quantum dots, confirming precise gate manipulability in ZnO quantum dot systems.
Furthermore, we have observed the QCA effect, which cannot be observed in single or double quantum dots.
These achievements progress the utilization of the unique material properties of ZnO for quantum applications, while also providing an interesting platform for exploring fundamental quantum phenomena.
The controllable multiple quantum dots in ZnO demonstrated in this study will open new possibilities for both scalable spin qubits and the investigation of fundamental physics in this promising material system.

\subsection{Notes}
The authors declare no competing financial interests.

\begin{acknowledgement}

The authors thank M. Takeuchi, A. Kurita, RIEC Fundamental Technology Center, and the Laboratory for Nanoelectronics and Spintronics for technical support.
Part of this work was supported by
MEXT Leading Initiative for Excellent Young Researchers, 
Grants-in-Aid for Scientific Research (21K18592, 22H04958, 23H01789, 23H04490), 
FRiD Tohoku University, 
and "Advanced Research Infrastructure for Materials and Nanotechnology in Japan (ARIM)" of the Ministry of Education, Culture, Sports, Science and Technology (MEXT) (Proposal Number JPMXP1224NM5072).
AIMR and MANA are supported by World Premier International Research Center Initiative (WPI), MEXT, Japan.

\end{acknowledgement}

\bibliography{reference}

\end{document}